\documentclass{aastex63}

\usepackage[T1]{fontenc}
\usepackage{wasysym}
\usepackage{graphicx}
\usepackage{makecell}
\usepackage{multirow}
\usepackage{tabularx}
\usepackage{soul}
\submitjournal{PSJ}

\shorttitle{Wet accretion of inner solar system bodies}
\shortauthors{Jin et al.}
\graphicspath{{./}{figures/}}

\begin{document}

\title{New evidence for wet accretion of inner solar system planetesimals from meteorites Chelyabinsk and Benenitra}

\correspondingauthor{Ziliang Jin}
\email{zljin@must.edu.mo}

\author[0000-0002-1852-9362]{Ziliang Jin}
\affiliation{School of Earth and Space Exploration \\
Arizona State University \\
Tempe, AZ 85287, USA}
\affiliation{State Key Laboratory of Lunar and Planetary Sciences \\
Macau University of Science and Technology\\ Taipa, Macau, China}

\author[0000-0002-7978-6370]{Maitrayee Bose}
\affiliation{School of Earth and Space Exploration \\
Arizona State University \\
Tempe, AZ 85287, USA}

\author[0000-0002-3286-7683]{Tim Lichtenberg}
\affiliation{Atmospheric, Oceanic and Planetary Physics \\ Department of Physics \\ University of Oxford \\ United Kingdom}

\author[0000-0002-1078-9493]{Gijs D. Mulders}
\affil{Facultad de Ingenier\'{i}a y Ciencias,\\ 
Universidad Adolfo Ib\'{a}\~{n}ez,\\
Pe\~{n}alol\'{e}n, Santiago, Chile}
\affil{Millennium Institute for Astrophysics,\\
Chile}

\begin{abstract}
We investigated the hydrogen isotopic compositions and water contents of pyroxenes in two recent ordinary chondrite falls, namely, Chelyabinsk (2013 fall) and Benenitra (2018 fall), and compared them to three ordinary chondrite Antarctic finds, namely Graves Nunataks GRA 06179, Larkman Nunatak LAR 12241, and Dominion Range DOM 10035. The pyroxene minerals in Benenitra and Chelyabinsk are hydrated ($\sim$0.018–0.087 wt.$\%$ H$_2$O) and show D-poor isotopic signatures ($\delta$D$_{SMOW}$ from -444$\permil$ to -49$\permil$). On the contrary, the ordinary chondrite finds exhibit evidence of terrestrial contamination with elevated water contents ($\sim$0.039–0.174 wt.$\%$) and $\delta$D$_{SMOW}$ values (from -199$\permil$ to -14$\permil$). We evaluated several small parent body processes that are likely to alter the measured compositions in Benenitra and Chelyabinsk, and inferred that water-loss in S-type planetesimals is minimal during thermal metamorphism. Benenitra and Chelyabinsk hydrogen compositions reflect a mixed component of D-poor nebular hydrogen and water from the D-rich mesostases. 45-95$\%$ of water in the minerals characterized by low $\delta$D$_{SMOW}$ values was contributed by nebular hydrogen. S-type asteroids dominantly composed of nominally anhydrous minerals can hold 254–518 ppm of water. Addition of a nebular water component to nominally dry inner Solar System bodies during accretion suggests a reduced need of volatile delivery to the terrestrial planets during late accretion. 

\end{abstract}


\section{Introduction} 
\label{sec:1}

 In the inner solar system, water is present in the terrestrial planets, as evident from the vast amounts of water that cover Earth’s surface and contained in its deep interior \citep[e.g.,][]{peslier2017water}, the subglacial liquid water detected in the polar region of Mars \citep[e.g.,][]{orosei2018radar}, the surface water ice in permanently shadowed regions of Mercury \citep[e.g.,][]{lawrence2013evidence}, and the water vapor measured in Venus’s atmosphere \citep[e.g.,][]{cottini2012water}. However, in the standard picture of planetary accretion, the location of the water snowline outside of the terrestrial planet formation region necessitates the addition of water-rich materials from further-out regions \citep{RaymondIzidoro2017}. Thus, the processes and timing of acquisition of water by the terrestrial planets is a long-standing question in planetary science, and no consensus has been reached yet. 

Several scenarios have been proposed to explain the source and delivery of water to the inner solar system based on the studies of hydrogen isotopic composition of planetary materials. The most popular scenario argues that water to Earth was delivered by the water-rich objects from the outer solar system, e.g., C-type asteroids or comets during different periods of planetary formation \citep[e.g.,][]{alexander2012provenances,sarafian2014early}. This scenario is based on the assumption that the building blocks of terrestrial planets were dry. However, D/H ratios of comets measured remotely do not well match that of Earth’s water, although one of the Jupiter-family comets 103P/Hartley 2 has comparable water to Earth \citep[e.g.,][]{hutsemekers2008oh,altwegg201567p,hartogh2011ocean}. In addition, although D/H ratios of CM and CI chondrites are comparable with terrestrial rocks, it is unlikely that C-type materials delivered a significant fraction of water to Earth. This is because in addition to water, C-type asteroids would have brought significant amounts of siderophile elements, e.g., Os and Ru, and these elements show distinct isotopic compositions from terrestrial materials \citep{meisel1996osmium,FischerGoedde2017}. It has been suggested that Earth should have incorporated more than 70$\%$ of its hydrogen prior to the late chondritic veneer \citep{HALLIDAY2013146}. Similarly, bulk abundance ratios of atmophile elements in Earth \citep{Hirschmann2016} and Venus' present-day atmosphere \citep{Gillmann2020} provide evidence against an overly volatile-rich composition of late-accreted material to the terrestrial planet region. Instead, this supports the addition of an early, indigeneous source of water to the inner solar system planets. 

Ingassing of nebular hydrogen into the magma ocean on a differentiating rocky planet have been invoked to explain the addition of hydrogen into Earth's interior because the solar nebular gas characterized by a low D/H ratio (2.1 × 10$^{-5}$, \citealp{geiss1998abundances}) was likely present during Earth's initial growth stages \citep{ikoma2006constraints,sharp2017nebular}. However, the solubility of hydrogen molecules in silicate melts is dependent on the atmosphere pressure above the surface of the magma ocean \citep{e48e6466417f426bb4f6f91830fd0eef}, which is not well constrained, and hence brings huge uncertainties on the amount of dissolved hydrogen. For instance, under a redox condition corresponding to the iron-w\"ustite buffer, solubility of H in the magma ocean rises from 10 to 40 ppm (equivalent to 90-360 ppm water) as the atmosphere pressure increases from 0.1 to 1 bar \citep{e48e6466417f426bb4f6f91830fd0eef}. Furthermore, the accretion of gaseous hydrogen envelopes tested through 1D gas accretion models \citep{coleman2019pebbles} show that Earth-size planets could only accrete about $1\%$ of their total mass in gas during its growth period, and the ingassing process became particularly slow during the post solid accretion phase. The possibility of the accreted envelope to be stripped away via irradiation by the central star is high, especially closer to the star \citep{Lammer2018AAR,Owen2019AREPS}. More importantly, the ingassing scenario is not an efficient process for small planetesimals ($<$250 km). The recently measured hydrogen isotopic compositions of the planetesimal population, including Itokawa regolith minerals and enstatite chondrites indicate that undifferentiated inner solar system small bodies contain a significant amount of D-depleted water \citep{jin2019new,piani2020earth}. Therefore, rather than the dissolved hydrogen in the magma ocean, the undifferentiated small bodies containing hydrogen (in various forms) are a more likely source of water for terrestrial planets.

In order to determine the source of water in the inner solar system bodies and evaluate the processes by which terrestrial planets can acquire water through planetary accretion, we investigated the D/H ratios and water contents of silicate minerals from two recent ordinary chondrite (OC) falls and three Antarctic OC finds. In light of these analytical results, we evaluated the parent body processes that might have modified the measured hydrogen signatures. We report high water contents and variable D/H ratios of the measured minerals; several grains in Chelyabinsk and Benenitra OC falls are characterized by D-poor signatures, indicating the wet accretion of chondritic planetesimals in the inner solar system. As a consequence, the high water content of planetary embryos and planets formed through planetary accretion is well-justified.

\section{Samples and analytical methods} \label{sec:2}

\subsection{Sample description and preparation} \label{subsec:2.1}
Two OC falls are LL5 Chelyabinsk and L6 Benenitra. Chelyabinsk fell in Chelyabinskaya district, Russia in February 2013. The fragments were collected immediately after its fall. Benenitra was collected in town of Benenitra, Madagascar right after its fall on 27th July 2018. Three OC finds, namely, LL4 Graves Nunataks 06179 (GRA 06179), LL5 Larkman Nunatak 12241 (LAR 12241), and L6 Dominion Range 10035 (DOM 10035) are Antarctic meteorites and their weathering scales are categorized as A/B, A, and B, respectively. 

Fresh fragments of the OC falls were acquired from the Field Museum, Chicago. These samples were polished with no hydrogen-bearing lubricants (‘dry polish’), mounted in 1-inch indium metal, and placed in the oven at 45 $^\circ$C for 1 week, before placing in the ultra-high vacuum chambers in the NanoSIMS 50L at Arizona State University (ASU). Because we wanted to measure hydrogen in the polished samples, addition of water or water-based organic solvents was completely avoided during the 'dry polish'. The polishing materials included diamond films with grain sizes of 6 $\mu$m, 3 $\mu$m, 1 $\mu$m, and 0.5 $\mu$m. Three OC finds were prepared as thin sections by Johnson Space Center.

\subsection{Standard description and preparation} \label{subsec:2.2}

The standards used to calibrate the water contents during NanoSIMS measurements were ALV-519 (basaltic glass, 1700 ± 43 ppm H$_2$O, 1$\sigma$, \citealp{SaalA.E2013HIiL}, sessions 1–3), PMR-53 (clinopyroxene, 268 ± 8 ppm H$_2$O, 1$\sigma$, \citealp{RossmanGeorgeR1995Qaot}, sessions 2-3), KH03-27 (orthopyroxene, 367 ± 49 ppm H$_2$O, 1$\sigma$, \citealp{KumamotoKathrynM2017NSrm}, session 1) and 116610-18 (orthopyroxene, 119 ± 11 ppm H$_2$O, 1$\sigma$, \citealp{KumamotoKathrynM2017NSrm}, used in sessions 1–3). ALV-519 ($\delta$D$_{SMOW}$ = -72$\permil$, \citealp{SaalA.E2013HIiL}) and PMR-53 ($\delta$D$_{SMOW}$ = -115$\permil$, \citealp{BellDavidR2000Tico}) were used to calibrate the D/H ratios of the minerals. The San Carlos olivine was used as the blank to monitor the background during NanoSIMS analyses.

The standards used during the IMS 6f session were ALV-519, 116610-18, and PMR-53. San Carlos olivine was used as the blank to calculate the hydrogen background.

Small chips of the acquired standards were first mounted in epoxy (EpoThin 2 Resin by Buehler) and dry-polished flat and smooth using sandpaper and diamond films. Then we used pure acetone to partially dissolve the epoxy stub and pick out the polished standards. Subsequently, the polished standards were dried in the oven for several weeks, followed by mounting into 10-mm aluminum rings filled with indium (99.9$\%$ pure). In each NanoSIMS session, the meteorite and standard mounts were placed in the same holder so that they can be measured under identical conditions.

\begin{figure}[ht]
    \centering
    \includegraphics[scale=0.3]{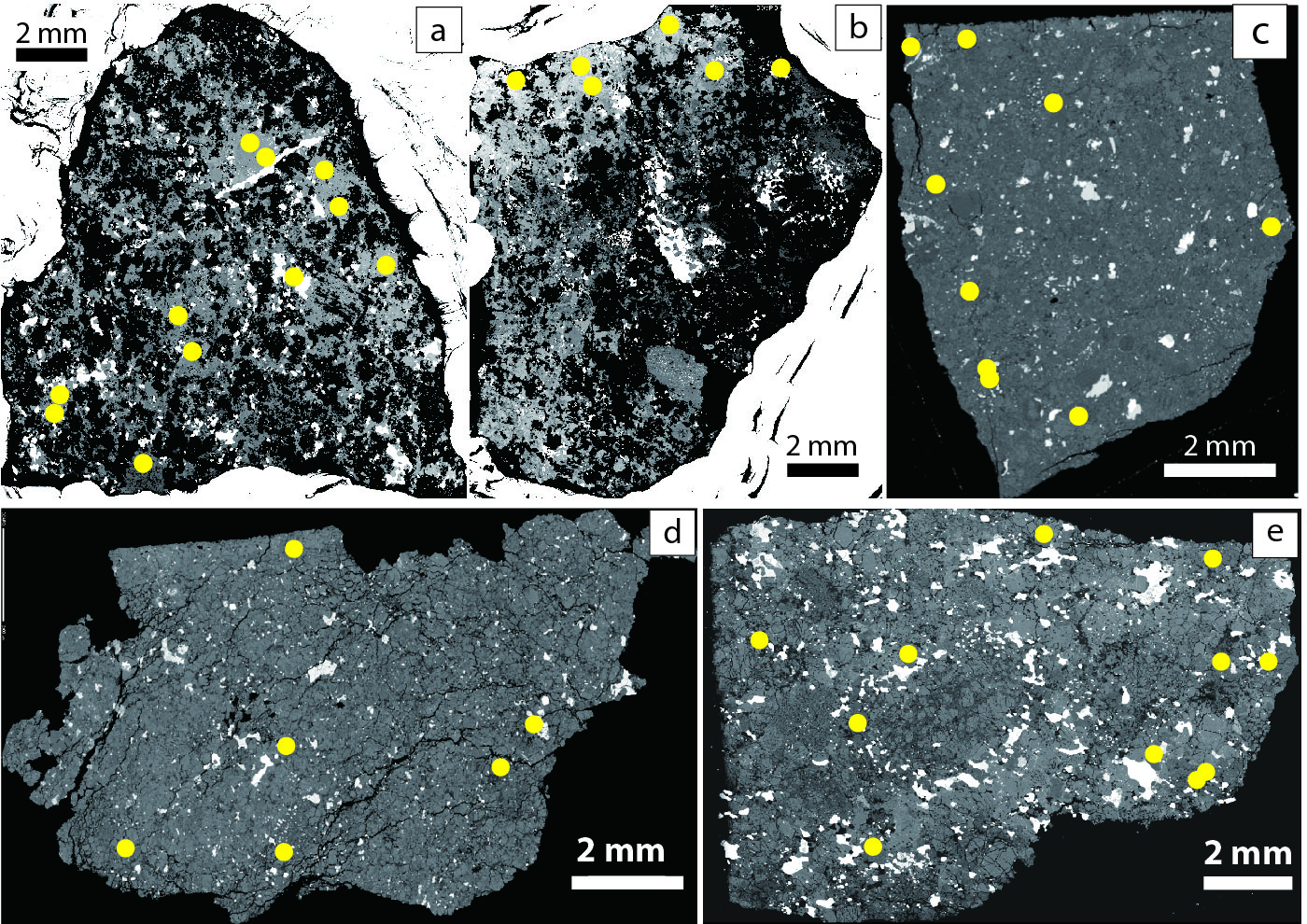}
    \caption{\textbf{Backscattered electron images of ordinary chondrites}. The measured OCs are (a) Benenitra, (b) Chelyabinsk, (c) LL4 GRA 06179, (d) LL5 LAR 12241, and (e) L6 DOM 10035. The yellow dots show the analyzed spots with the SIMS instruments.}
    \label{fig:1}
\end{figure}

\subsection{Backscattered electron (BSE) images and elemental mapping} \label{subsec:2.3} 
Backscattered electron images (Fig. \ref{fig:1}) and X-ray intensity maps (Si, Ti, Al, Fe, Mg, Ca, Na, and K) of the meteorites were obtained using a JEOL JXA-8530F electron microprobe at ASU with an accelerating voltage of 15 keV, a beam current of 60 nA, a beam diameter of 20 $\mu$m, and a dwell time of 30 $\mu$s/pixel. The elemental maps were used to estimate the volume fractions of mineral phases in the analyzed meteorites, and to identify and select orthopyroxene minerals that were later measured for water contents and D/H ratios by the ion microprobes. The parent bodies of these type 5 and 6 meteorites are supposed to have experienced thermal metamorphism characterized by peak temperatures up to 1100 K \citep{GailHans-Peter2019Thmo}. The original chondritic silicate minerals, namely, olivine and pyroxene, crystallized from the precursor melts at temperatures of $\>$1400 K would have survived under the high temperatures, whereas the chondrule metastasis and fine-grained matrix components recrystallized and formed new mineral phases, e.g., feldspar. We selected pyroxene grains (20-100 $\mu$m) to measure because they are more resistant to thermal metamorphism than olivines in chondrules \citep{huss2006thermal}.

\subsection{Secondary ion mass spectrometry (SIMS)} \label{subsec:2.4} 
\textit{Cameca NanoSIMS 50L.} D/H ratios and H$_2$O concentrations of the standards and orthopyroxenes from Chelyabinsk, Benenitra, GRA 06179, and DOM 10035 were measured by the Cameca NanoSIMS 50L. The NanoSIMS measurements were done in 3 sessions: Session 1 - GRA 06179 and DOM 10035; Session 2 - Benenitra; Session 3 - Chelyabinsk. A 220 pA Cs$^+$ primary beam with an approximate beam size of $<$2 $\mu$m (D1-3 aperture with diameter of 200 $\mu$m) was rastered on a 15 × 15 $\mu$m$^2$ surface on the grains. Electronic gating was used to collect the secondary ion signal from the internal 25\% of the rastered  area. The primary beam has an impact energy of 16 keV. Prior to the data collection, the sample surface was presputtered for $\sim$10 mins in order to remove the surface contamination. $^{1}$H$^-$, $^{2}$D$^-$, $^{12}$C$^-$, and $^{18}$O$^-$ were detected simultaneously with a 50-$\mu$m entrance slit (ES1) and no aperture slit. The mass resolving power (MRP) for the spectrometer was $\sim$1250 (Cameca mass resolving power). Each measurement contained 200 cycles. During the analyses, the counting time for each pixel was 300 $\mu$s/pixel. To compensate for the charging of the sample surface, the electron gun ($\sim$80 nA) was employed. The pressure of the analysis chamber was better than 2 × 10$^{-10}$ torr.

\textit{Cameca IMS 6f.} The measurements of water contents and D/H ratios in the pyroxenes from LAR 12241 (Session 4) were performed on the Cameca IMS 6f at ASU. A 10–13 nA Cs$^+$ ion beam accelerated with an energy of 15 keV was rastered on a 35 × 35 $\mu$m$^2$ surface area. The size of the primary beam was $\sim$25 $\mu$m. The area of interest was presputtered for 2 minutes prior to analysis. The field aperture was set to 15 $\mu$m diameter in order to reduce the crater edge effects. For each spot, 50 cycles measuring H$^-$ and D$^-$ were collected by an electron multiplier with counting times of 1 and 10 seconds, respectively.  At the end of each measurement, $^{16}$O$^-$ was measured using a Faraday cup. The mass resolving power during operation was $\sim$800. The electron gun with a curent of $\sim$1 $\mu$A was used to compensate for surface charge during the measurements. The vacuum in the analytical chamber was $\sim$3 × 10$^{-9}$ torr during analyses. 

\subsection{Wavelength-dispersive X-ray spectroscopy (WDS)} \label{subsec:2.5} 
The orthopyroxenes that have been analyzed using SIMS instruments were then measured using wavelength-dispersive X-ray spectroscopy. Prior to the WDS analysis, we checked the craters left by the SIMS measurements and only selected the areas with clean and flat craters to conduct the WDS measurements (Fig. \ref{fig:2}). The analyzed area on each mineral grain was close to the SIMS crater. Analytical conditions were 15 keV acceleration voltage, 15 nA beam current, 1 $\mu$m beam diameter, and counting times on the peaks and backgrounds of 40 s for Na and K, and 60 s for Si, Ti, Al, Cr, Ca, Mg, Mn, and Fe. Well-characterized, natural and synthetic standards (K-hornblende, Ca-wollastonite, Al, Na-augite, Si, Mg-enstatite, Ti-rutile, Cr-chromite, Fe-fayalite) were used for calibration and to confirm the analytical precision. Detection limits were 0.02 wt.$\%$ for K$_2$O, CaO, Al$_2$O$_3$, SiO$_2$, Na$_2$O, and MgO, 0.03 wt.$\%$ for TiO$_2$ and Cr$_2$O$_3$, and 0.06 wt.$\%$ for MnO and FeO; the relative error of the analyses was better than 0.3$\%$ for Si and Mg, $\sim$1$\%$ for FeO, $\sim$10$\%$ for Ca, Mn, and Cr, and larger than 20$\%$ for K, Al, Na, and Ti.

\section{Results} \label{sec:3}
\subsection{Petrologic characterization}\label{subsec:3.1}
\begin{figure}[ht]
    \centering
    \includegraphics[scale=0.3]{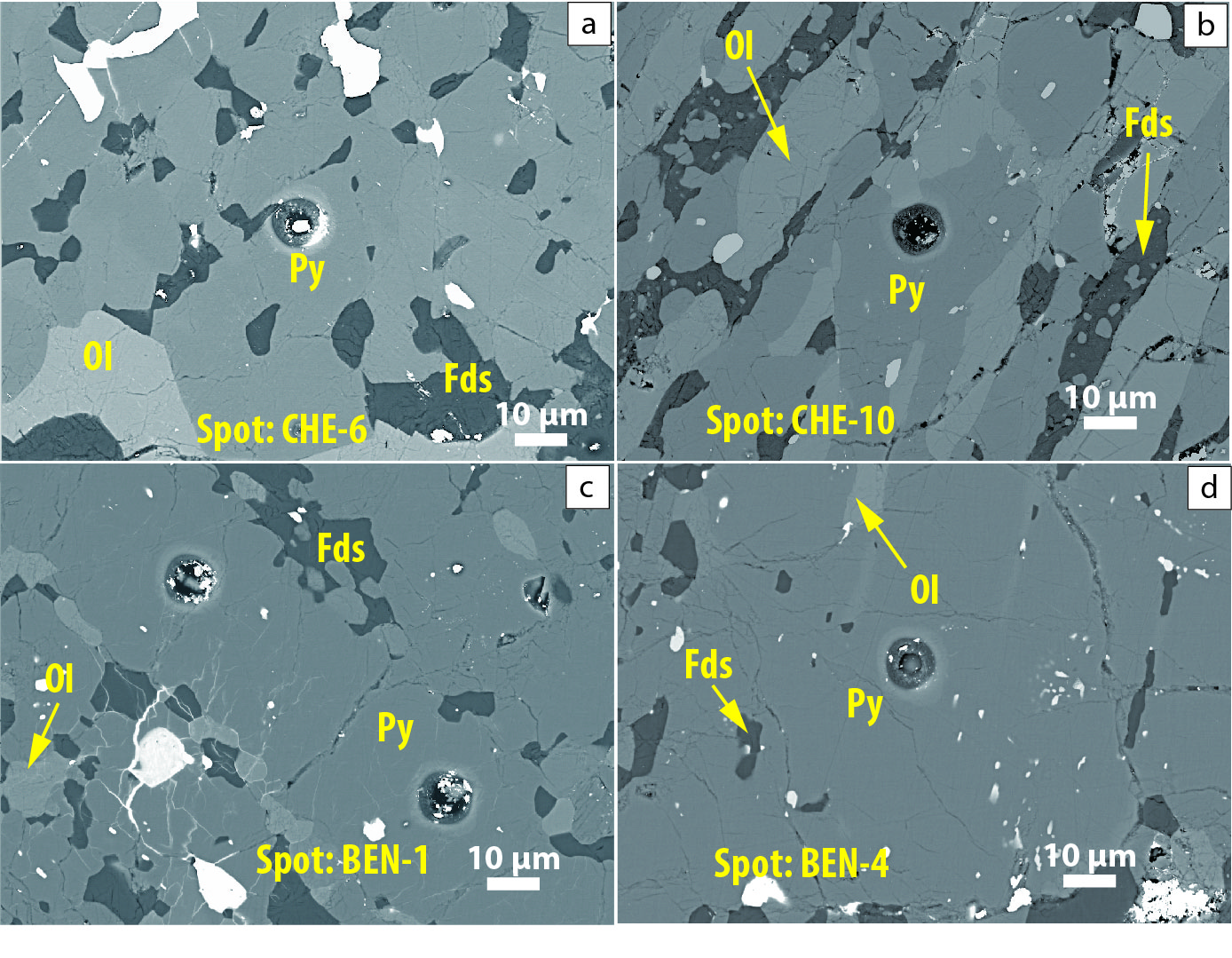}
    \caption{\textbf{Representative backscattered electron images of Chelyabinsk (a-b) and Benenitra (c-d)}. The major silicate minerals in both ordinary chondrites are olivine (Ol) and pyroxene (Py). Feldspars (Fds) occur in the interstitial areas. The small craters were generated after NanoSIMS measurements.}
    \label{fig:2}
\end{figure}

We estimated the mineral proportions of two meteorite falls, namely, Chelyabinsk and Benenitra, based on the X-ray elemental maps. The dominant silicate minerals in Chelyabinsk are olivine (60$\pm$5 vol.$\%$) and low-Ca pyroxene (30$\pm$5 vol.$\%$), consistent with previous reports \citep{ozawa2014jadeite,kaeter2018chelyabinsk}. A smaller fraction of feldspars ($<$10 vol.$\%$) and Fe-Ni metal ($<$5 vol.$\%$) also occur in the interstitial areas (Fig. \ref{fig:2}a and b). The fragment of Benenitra consists of olivine (55$\pm$5 vol.$\%$)and pyroxene (30$\pm$5 vol.$\%$), interstitial feldspars ($\sim$10 vol.$\%$) and Fe-Ni metals (Fig. \ref{fig:2}c and d). The original matrix and chondrule mesostasis components in the two meteorites are absent because they have experienced extensive thermal metamorphism.
 
\subsection{Water contents} \label{subsec:3.2}
\begin{figure}[ht]
    \centering
    \includegraphics[scale=0.17]{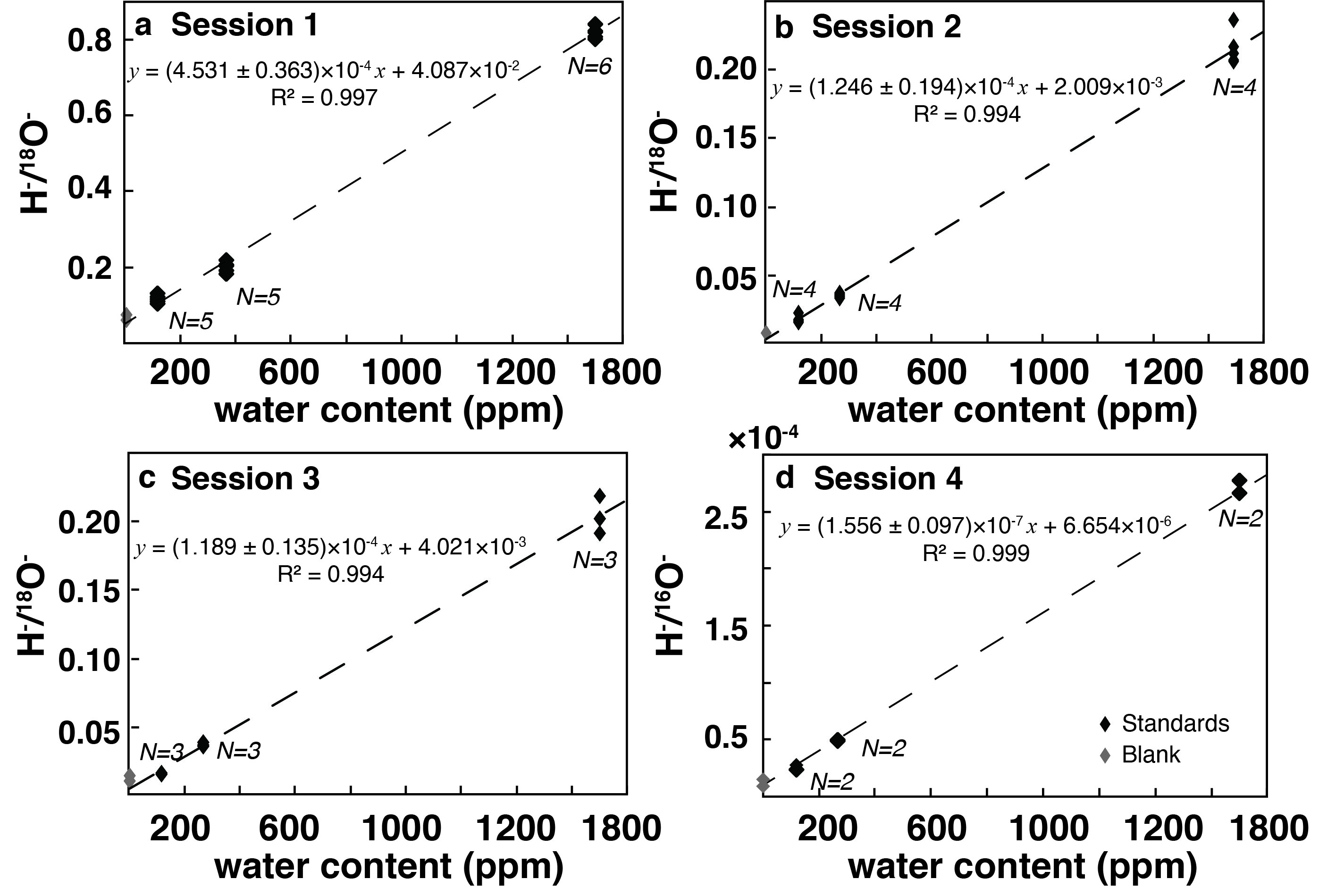} 
    \caption{\textbf{Calibrations of water contents and hydrogen isotope compositions using terrestrial mineral and glass standards}. The measured H/O ion ratios vs. H$_2$O contents derived through bulk measurements are plotted. The trendlines shown in the diagrams were used to estimate the water contents of the measured pyroxenes in Sessions 1–4 (a-d). The standards are: ALV-519 (basaltic glass, 1700 ± 43 ppm H$_2$O), PMR-53 (clinopyroxene, 268 ± 8 ppm H$_2$O), KH03-27 (orthopyroxene, 367 ± 49 ppm H$_2$O) and 116610-18 (orthopyroxene, 119 ± 11 ppm H$_2$O). Blank monitor: San Carlos olivine.}
    \label{fig:3}
\end{figure}

In nominally anhydrous minerals (NAMs), hydrogen generally occurs in the defects of the crystal, because normal lattice site cannot be occupied by hydrogen. For instance, the vacancies of Mg and Si in NAMs can be occupied by hydrogen via substitutions 2H$^+$ $\rightarrow$ Mg$^{2+}$ and H$^+$+Al$^{3+}$ $\rightarrow$ Si$^{4+}$ \citep{bell1992water,hauri2006partitioning}. These hydrogen ions preferentially combine with the adjacent oxygen ions and form structurally bound O-H in the crystal lattice. The O-H bonds in the crystal act as the proxy for water. During SIMS measurements, the O-H bonds are broken by the primary ion beam. The released secondary H$^+$ ions were subsequently collected by the detector on the instrument. Accordingly, we estimated the water contents of the measured pyroxenes in both OC falls and finds by using the H$^-$/O$^-$ versus H$_2$O calibration line. We first plotted the measured H$^-$/O$^-$ ratios of the standards versus their reported water contents in previous literature (Fig. \ref{fig:3}). Then we used the slopes and the intercept values of the calibration lines and the H$^-$/O$^-$ ratios of the measured pyroxenes in OC samples to calculate their water contents. The low H$^-$/O$^-$ values of San Carlos olivine correspond to background water of 65 ppm, 34 ppm, 67 ppm, and $<$15 ppm for sessions 1–4. The uncertainties on the water contents in the samples were obtained by propagating the uncertainties on the H$^-$/O$^-$ ratios and slopes of the calibration trendlines.

The water contents of orthopyroxenes from OC finds and falls are listed in Table \ref{tab:1}. The minerals from OC finds have water contents of 458–1807 parts per million (ppm) weight. In OC falls, the measured orthopyroxenes have water contents of 210–902 ppm. After subtracting the water from the background, the water contents of pyroxenes in OC finds are 393–1742 ppm (median = 844$\pm$173 ppm) and in OC falls contain 176–868 ppm water (median = 558$\pm$116 ppm). The water contents of pyroxene minerals in OC falls (176-868 ppm) are comparable to those of the returned Itokawa pyroxene minerals \citep[278-988 ppm,][]{jin2019new,chan2021organic}, yet the upper limit (868 ppm) is lower than that observed in several pyroxenes from the Type 3 OC Bishunpur \citep[1796 ppm,][]{stephant2017water}. In contrast, our results are all higher than the reported water content of one pyroxene grain from the Type 3.05 OC QUE 97008 \citep[12 ppm,][]{shimizu2021highly}.

\subsection{Hydrogen isotopes} \label{subsec:3.3}

During SIMS measurements, the instrumental mass fractionation (IMF) as a result of the sputtering process, transmission efficiency, and ion detection efficiency can cause systematic bias in the data, which must be corrected. The IMF is defined as
\begin{equation}\label{eq:1}
   IMF = \frac{(D/H)_{MEAS}-(D/H)_{TRUE}}{(D/H)_{TRUE}},
\end{equation}
where (D/H)$_{MEAS}$ is the measured ratio and (D/H)$_{TRUE}$ is the real isotopic ratio of the sample. The real and measured D/H ratios of ALV-519 [(D/H)$_{TRUE}$ = 1.4454 × 10$^{-4}$] and PMR-53  [(D/H)$_{TRUE}$ = 1.3784 × 10$^{-4}$] were used to calculate the IMFs during the analyses. In session 1, the average IMF of ALV-519 is 0.490 ± 0.053 ( N = 6, 1$\sigma$). In sessions 2 and 3, the IMFs of ALV-519 and PMR-53 are similar within errors in each session. We then employed the average IMF of the utilized references (ALV-519 and PMR-53), which are -0.041 ± 0.039 (N = 8, 1$\sigma$) for session 2 and -0.078 ± 0.021 (N = 6, 1$\sigma$) for session 3. The IMF for IMS 6f measurements is -0.013 ± 0.019 (Session 4, N = 4, 1$\sigma$).

Using these IMF numbers, we obtained the true D/H ratios of measured minerals. Then we made the background correction for the D/H ratios [(D/H)$_{CORR}$] by subtracting the portion of D and H contributed by the background hydrogen. These corrected D/H ratios were normalized to Standard Mean Ocean Water (SMOW, D/H = 1.5576 × 10$^{-4}$), via
\begin{equation}\label{eq:2}
    \delta D_{SMOW} (\permil) = (\frac{(D/H)_{CORR}}{(D/H)_{SMOW}}-1)*1000.
\end{equation}
The D/H ratios and associated errors of orthopyroxenes from OC finds and falls are listed in Table \ref{tab:1}. Uncertainties on the D/H ratios and the IMF were propagated to calculate the uncertainties on the $\delta$D$_{SMOW}$ values. The $\delta$D$_{SMOW}$ values are plotted against the measured water contents in Fig. \ref{fig:3}. The $\delta$D$_{SMOW}$ values of measured minerals from OC finds vary from -199$\permil$  to -14$\permil$ (Median = -129$\pm$24$\permil$). The measured pyroxenes from OC falls have more negative $\delta$D$_{SMOW}$ values varying from -444$\permil$ to -49$\permil$, with a median value of -263$\pm$63$\permil$.

\begin{figure}[ht]
    \centering
    \includegraphics[scale=0.2]{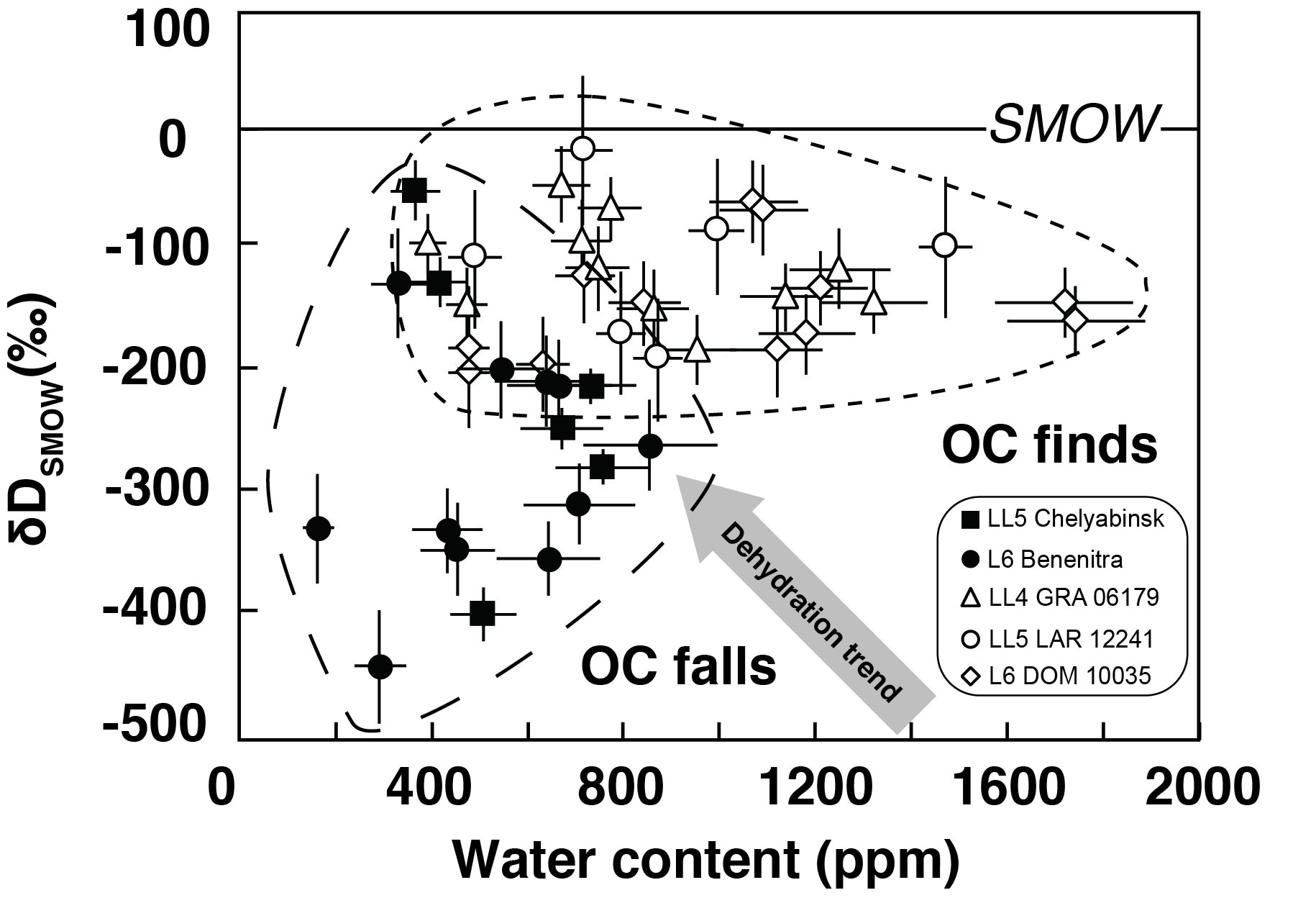}
    \caption{\textbf{Water contents (ppm) and $\delta$D$_{SMOW}$ values ($\permil$) of pyroxenes from ordinary chondrite (OC) falls and finds}. Minerals from OC falls have distinguishable and much lower $\delta$D$_{SMOW}$ values (as low as -444$\permil$ observed in Benenitra) and water contents than those of minerals from OC finds. No clear dehydration trend is observed. Errors are 1 standard deviation. SMOW: standard mean ocean water.}
    \label{fig:4}
\end{figure}

\subsection{Major element abundances} \label{subsec:3.4}
The measured minerals from the OC finds have
similar major element compositions. They have MgO contents of 24.30–30.43 wt.$\%$, FeO contents of 12.30–18.02 wt.$\%$. The Al$_{2}$O$_{3}$ contents of measured minerals in OC finds mostly fall in a range of 0.10–0.32 wt.$\%$, except one grain has 2.24 wt.$\%$ Al$_{2}$O$_{3}$ (Table \ref{tab:2}). The three end members [Wollastonite (Wo)-Enstatite (En)-Ferrosilite (Fs)] composition range is Wo$_{0.4-3.9}$ En$_{70.4-81.0}$ Fs$_{18.4-27.6}$. The measured minerals from these OC falls have MgO contents of 27.36–29.40 wt.$\%$, FeO contents of 13.58–15.91 wt.$\%$, and Al$_{2}$O$_{3}$ contents of 0.08–0.21 wt.$\%$ (Table \ref{tab:2}). Their Wo-En-Fs compositions are similar to those of OC finds and have a range of Wo$_{0.9-2.0}$ En$_{74.5-78.1}$ Fs$_{20.5-24.1}$.

\begin{figure}[ht]
    \centering
    \includegraphics[scale=0.2]{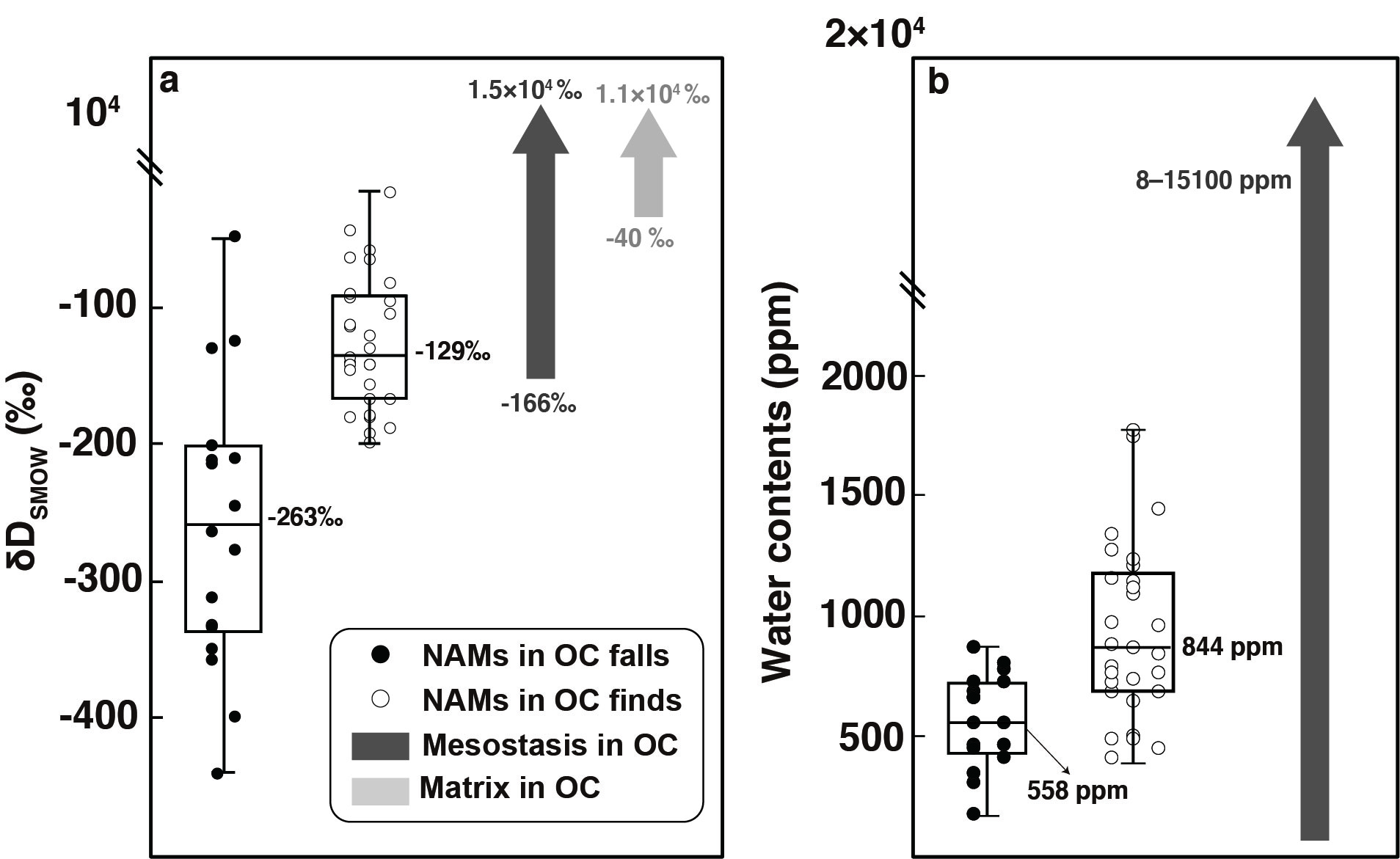}
    \caption{\textbf{Box-and-whisker plots of measured (a) $\delta$D$_{SMOW}$ values and (b) water contents}. The interquartile range covering values between the first and third quartile of a dataset is indicated by the box. The upper and lower whiskers represent dataset numbers outside the interquartile range. Solid line in each box indicates the median value of a dataset. The interquartile ranges of water contents and $\delta$D$_{SMOW}$ values of minerals in OC finds are higher than those of minerals in OC falls. Interstitial components of chondrites, e.g., mesostasis and matrix, generally have large variations of D/H ratios water contents. Data of the interstitial components are from \cite{PianiLaurette2015MDhi,stephant2017water,shimizu2021highly}.}
    \label{fig:5}
\end{figure}

\begin{table}[ht]
    \caption{Hydrogen isotopic compositions and water contents of pyroxene minerals in ordinary chondrite falls and finds.}
\begin{tabular}[c]{ccccccc}
\hline
\hline
Sample      & D/H$^*$      & 1$\sigma$       & $\delta$D$_{SMOW}$ & 1$\sigma$ & Water content (ppm)$^*$ & 1$\sigma$ \\ \hline
\multicolumn{7}{c}{NanoSIMS Session 1}                                      \\ \hline
GRA06179-1  & 1.46 $\times$ 10$^{-4}$ & 7.85 $\times$ 10$^{-6}$ & -63          & 34      & 773   & 67     \\
GRA06179-2  & 1.38 $\times$ 10$^{-4}$ & 6.80 $\times$ 10$^{-6}$ & -112         & 29      & 746   & 65     \\
GRA06179-3  & 1.42 $\times$ 10$^{-4}$ & 6.99 $\times$ 10$^{-6}$ & -89          & 30      & 712   & 62     \\
GRA06179-5  & 1.41 $\times$ 10$^{-4}$ & 9.67 $\times$ 10$^{-6}$ & -92          & 42      & 393   & 37     \\
GRA06179-6  & 1.49 $\times$ 10$^{-4}$ & 7.74 $\times$ 10$^{-6}$ & -42          & 33      & 670   & 59     \\
GRA06179-7  & 1.28 $\times$ 10$^{-4}$ & 6.75 $\times$ 10$^{-6}$ & -180         & 29      & 954  & 82     \\
GRA06179-8  & 1.35 $\times$ 10$^{-4}$ & 6.85 $\times$ 10$^{-6}$ & -136         & 29      & 1139  & 97     \\
GRA06179-10 & 1.38 $\times$ 10$^{-4}$ & 5.62 $\times$ 10$^{-6}$ & -113         & 24      & 1250  & 105    \\
GRA06179-11 & 1.33 $\times$ 10$^{-4}$ & 6.74 $\times$ 10$^{-6}$ & -145         & 29      & 862   & 74     \\
GRA06179-12 & 1.34 $\times$ 10$^{-4}$ & 8.82 $\times$ 10$^{-6}$ & -142         & 38      & 475   & 43     \\
DOM10035-1  & 1.28 $\times$ 10$^{-4}$ & 8.96 $\times$ 10$^{-6}$ & -180         & 38      & 1118  & 95     \\
DOM10035-2  & 1.47 $\times$ 10$^{-4}$ & 8.04 $\times$ 10$^{-6}$ & -57          & 35      & 1070  & 91     \\
DOM10035-3  & 1.34 $\times$ 10$^{-4}$ & 8.18 $\times$ 10$^{-6}$ & -141         & 35      & 844   & 73     \\
DOM10035-4  & 1.26 $\times$ 10$^{-4}$ & 9.08 $\times$ 10$^{-6}$ & -192         & 39      & 632   & 56     \\
DOM10035-5  & 1.37 $\times$ 10$^{-4}$ & 8.91 $\times$ 10$^{-6}$ & -119         & 38      & 719   & 63     \\
DOM10035-7  & 1.34 $\times$ 10$^{-4}$ & 6.90 $\times$ 10$^{-6}$ & -140         & 30      & 1718  & 143    \\
DOM10035-11 & 1.32 $\times$ 10$^{-4}$ & 6.78 $\times$ 10$^{-6}$ & -156         & 29      & 1742  & 145    \\
DOM10035-12 & 1.25 $\times$ 10$^{-4}$ & 1.08 $\times$ 10$^{-5}$ & -199         & 47      & 479   & 44     \\
DOM10035-13 & 1.28 $\times$ 10$^{-4}$ & 1.22 $\times$ 10$^{-5}$ & -179         & 53      & 477   & 43     \\
DOM10035-14 & 1.46 $\times$ 10$^{-4}$ & 8.64 $\times$ 10$^{-6}$ & -64          & 37      & 1092  & 93     \\
DOM10035-15 & 1.36 $\times$ 10$^{-4}$ & 7.27 $\times$ 10$^{-6}$ & -129         & 31      & 1208  & 102    \\\hline
\multicolumn{7}{c}{NanoSIMS Session 2}                                      \\ \hline
Ben-1       & 1.23 $\times$ 10$^{-4}$ & 2.52 $\times$ 10$^{-6}$ & -210         & 36      & 653   & 106    \\
Ben-3       & 8.66 $\times$ 10$^{-5}$ & 5.84 $\times$ 10$^{-6}$ & -444         & 46      & 305   & 53     \\
Ben-4       & 1.04 $\times$ 10$^{-4}$ & 4.93 $\times$ 10$^{-6}$ & -331         & 44      & 176   & 33     \\
Ben-5       & 1.36 $\times$ 10$^{-4}$ & 3.94 $\times$ 10$^{-6}$ & -130         & 44      & 342   & 58     \\
Ben-6       & 1.01 $\times$ 10$^{-4}$ & 3.86 $\times$ 10$^{-6}$ & -349         & 37      & 465   & 77     \\
Ben-7       & 1.15 $\times$ 10$^{-4}$ & 2.94 $\times$ 10$^{-6}$ & -263         & 36      & 868   & 140    \\
Ben-8       & 1.04 $\times$ 10$^{-4}$ & 2.96 $\times$ 10$^{-6}$ & -333         & 34      & 446   & 74     \\
Ben-9       & 1.23 $\times$ 10$^{-4}$ & 2.65 $\times$ 10$^{-6}$ & -214         & 36      & 679   & 110    \\
Ben-10      & 1.25 $\times$ 10$^{-4}$ & 3.16 $\times$ 10$^{-6}$ & -201         & 39      & 558   & 92     \\
Ben-12      & 1.07 $\times$ 10$^{-4}$ & 2.34 $\times$ 10$^{-6}$ & -311         & 32      & 720   & 117    \\
Ben-13      & 1.00 $\times$ 10$^{-4}$ & 2.27 $\times$ 10$^{-6}$ & -356         & 31      & 657   & 107    \\ \hline
\multicolumn{7}{c}{NanoSIMS Session 3}                                      \\ \hline
Che-5       & 9.37 $\times$ 10$^{-5}$ & 2.22 $\times$ 10$^{-6}$ & -398         & 21      & 524   & 67     \\
Che-6       & 1.18 $\times$ 10$^{-4}$ & 2.35 $\times$ 10$^{-6}$ & -244         & 16      & 690   & 86     \\
Che-7       & 1.36 $\times$ 10$^{-4}$ & 2.76 $\times$ 10$^{-6}$ & -124         & 19      & 432   & 57     \\
Che-8       & 1.48 $\times$ 10$^{-4}$ & 3.37 $\times$ 10$^{-6}$ & -49          & 23      & 382   & 51     \\
Che-9       & 1.23 $\times$ 10$^{-4}$ & 2.04 $\times$ 10$^{-6}$ & -210         & 14      & 749   & 93     \\
Che-10      & 1.13 $\times$ 10$^{-4}$ & 1.97 $\times$ 10$^{-6}$ & -277         & 14      & 773   & 95     \\ 

 \hline
\multicolumn{7}{c}{IMS 6f Session 4}                                                  \\ \hline
LAR12141-2  & 1.30 $\times$ 10$^{-4}$ & 1.14 $\times$ 10$^{-6}$ & -167         & 50      & 794   & 51     \\
LAR12141-5  & 1.40 $\times$ 10$^{-4}$ & 1.83 $\times$ 10$^{-6}$ & -104         & 55      & 489   & 31     \\
LAR12141-6  & 1.54 $\times$ 10$^{-4}$ & 1.35 $\times$ 10$^{-6}$ & -14          & 59      & 717   & 46     \\
LAR12141-8  & 1.43 $\times$ 10$^{-4}$ & 1.41 $\times$ 10$^{-6}$ & -81          & 56      & 992  & 63     \\
LAR12141-10 & 1.27 $\times$ 10$^{-4}$ & 2.16 $\times$ 10$^{-6}$ & -187         & 51      & 871   & 55     \\
LAR12141-11 & 1.41 $\times$ 10$^{-4}$ & 1.96 $\times$ 10$^{-6}$ & -94          & 56      & 1468  & 93     \\
\hline
\hline
$^*$Background corrected.
\end{tabular}
\label{tab:1}
\end{table}

\begin{table}[ht]
\centering
\caption{Chemical compositions of the measured pyroxenes in the OC meteorites}
\begin{tabular}{cp{0.8cm}p{0.8cm}p{0.8cm}p{0.8cm}p{0.8cm}p{0.8cm}p{0.8cm}p{0.8cm}p{0.8cm}p{0.8cm}p{0.8cm}p{0.4cm}p{0.5cm}p{0.5cm}}
\hline
\hline
Spot & 
SiO$_2$ (wt.$\%$) &
  TiO$_2$  (wt.$\%$) &
  Al$_2$O$_3$ (wt.$\%$) &
  FeO (wt.$\%$) &
  MnO (wt.$\%$) &
  MgO (wt.$\%$) &
  CaO (wt.$\%$) &
  Na$_2$O (wt.$\%$) &
  K$_2$O (wt.$\%$) &
  Cr$_2$O$_3$ (wt.$\%$) &
  Total (wt.$\%$) &
  Wo ($\%$)&
  En ($\%$)&
  Fs ($\%$)\\
\hline
\multicolumn{15}{c}{OC finds}  \\
\hline
GRA06179-1  & 55.13 & 0.04 & 0.26 & 14.94 & 0.47 & 28.54 & 0.31 & bdl  & 0.00 & 0.26 & 99.94  & 0.6 & 76.8 & 22.6 \\
GRA06179-2  & 54.42 & 0.16 & 0.32 & 14.22 & 0.51 & 27.63 & 2.01 & 0.03 & 0.01 & 0.33 & 99.63  & 3.9 & 74.6 & 21.5 \\
GRA06179-3  & 55.29 & 0.01 & 0.13 & 14.21 & 0.48 & 29.23 & 0.29 & 0.02 & 0.00 & 0.19 & 99.86  & 0.6 & 78.1 & 21.3 \\
GRA06179-5  & 55.94 & 0.02 & 0.12 & 12.30 & 0.40 & 30.43 & 0.32 & 0.01 & 0.01 & 0.23 & 99.76  & 0.6 & 81.0 & 18.4 \\
GRA06179-6  & 54.00 & 0.11 & 2.24 & 15.50 & 0.41 & 24.30 & 1.87 & 0.11 & 0.05 & 0.36 & 98.95  & 3.9 & 70.8 & 25.3 \\
GRA06179-7  & 55.17 & 0.09 & 0.10 & 15.09 & 0.49 & 28.43 & 0.72 & 0.01 & 0.02 & 0.14 & 100.25 & 1.4 & 76.0 & 22.6 \\
GRA06179-8  & 55.60 & 0.02 & 0.14 & 13.74 & 0.37 & 29.69 & 0.20 & 0.00 & 0.01 & 0.18 & 99.93  & 0.4 & 79.1 & 20.5 \\
GRA06179-10 & 54.15 & 0.18 & 0.14 & 16.30 & 0.49 & 26.57 & 1.32 & 0.04 & 0.01 & 0.15 & 99.34  & 2.6 & 72.5 & 24.9 \\
GRA06179-11 & 54.19 & 0.18 & 0.11 & 16.07 & 0.53 & 26.73 & 1.32 & 0.02 & 0.00 & 0.12 & 99.27  & 2.6 & 72.8 & 24.6 \\
GRA06179-12 & 54.06 & 0.05 & 0.22 & 14.84 & 0.49 & 28.58 & 0.30 & 0.02 & 0.02 & 0.21 & 98.80  & 0.6 & 77.0 & 22.4 \\
Dom10035-1  & 55.39 & 0.12 & 0.12 & 13.66 & 0.48 & 28.99 & 0.94 & 0.02 & 0.03 & 0.15 & 99.90  & 1.8 & 77.7 & 20.5 \\
Dom10035-2  & 55.18 & 0.19 & 0.16 & 13.66 & 0.47 & 28.89 & 0.84 & 0.01 & 0.02 & 0.13 & 99.55  & 1.6 & 77.8 & 20.6 \\
Dom10035-3  & 54.95 & 0.15 & 0.15 & 13.85 & 0.50 & 28.49 & 0.77 & 0.04 & 0.02 & 0.10 & 99.01  & 1.5 & 77.4 & 21.1 \\
Dom10035-4  & 55.50 & 0.14 & 0.12 & 13.88 & 0.51 & 29.25 & 0.90 & 0.01 & 0.01 & 0.14 & 100.46 & 1.7 & 77.6 & 20.7 \\
Dom10035-5  & 55.46 & 0.19 & 0.17 & 13.99 & 0.51 & 28.71 & 1.12 & 0.02 & 0.00 & 0.15 & 100.34 & 2.2 & 76.8 & 21.0 \\
Dom10035-7  & 55.03 & 0.16 & 0.15 & 14.40 & 0.45 & 28.85 & 0.89 & 0.01 & 0.03 & 0.13 & 100.11 & 1.7 & 76.8 & 21.5 \\
Dom10035-11 & 55.49 & 0.18 & 0.18 & 13.94 & 0.49 & 29.02 & 0.85 & 0.01 & 0.02 & 0.10 & 100.27 & 1.6 & 77.5 & 20.9 \\
Dom10035-12 & 55.27 & 0.19 & 0.16 & 13.94 & 0.50 & 28.68 & 0.80 & bdl  & bdl  & 0.11 & 99.64  & 1.5 & 77.4 & 21.1 \\
Dom10035-13 & 55.32 & 0.18 & 0.16 & 13.40 & 0.54 & 28.77 & 1.07 & 0.03 & 0.01 & 0.15 & 99.62  & 2.1 & 77.6 & 20.3 \\
Dom10035-14 & 55.60 & 0.18 & 0.17 & 13.77 & 0.47 & 28.96 & 0.83 & 0.01 & 0.01 & 0.12 & 100.11 & 1.6 & 77.7 & 20.7 \\
Dom10035-15 & 55.52 & 0.20 & 0.18 & 13.81 & 0.45 & 28.76 & 1.00 & 0.02 & 0.00 & 0.19 & 100.13 & 1.9 & 77.3 & 20.8 \\
LAR12241-2  & 54.69 & 0.21 & 0.20 & 16.69 & 0.50 & 26.64 & 1.16 & 0.04 & bdl  & 0.17 & 100.30 & 2.3 & 72.3 & 25.4 \\
LAR12241-5  & 55.02 & 0.19 & 0.16 & 16.47 & 0.47 & 27.02 & 0.99 & 0.00 & bdl  & 0.16 & 100.45 & 1.9 & 73.1 & 25.0 \\
LAR12241-6  & 53.99 & 0.09 & 0.11 & 16.49 & 0.43 & 26.90 & 1.22 & 0.03 & 0.02 & 0.12 & 99.41  & 2.4 & 72.6 & 25.0 \\
LAR12241-8  & 53.47 & 0.17 & 0.18 & 18.02 & 0.43 & 25.74 & 0.99 & 0.03 & 0.02 & 0.11 & 99.16  & 1.9 & 70.4 & 27.6 \\
LAR12241-10 & 54.29 & 0.18 & 0.19 & 16.72 & 0.49 & 27.05 & 0.99 & 0.01 & 0.02 & 0.11 & 100.04 & 1.9 & 72.8 & 25.3 \\
LAR12241-11 & 54.59 & 0.21 & 0.16 & 16.76 & 0.49 & 26.97 & 1.15 & 0.02 & 0.02 & 0.14 & 100.52 & 2.2 & 72.5 & 25.3 \\
\hline
\multicolumn{15}{c}{OC falls} \\
\hline
BEN-1       & 55.57 & 0.18 & 0.15 & 13.99 & 0.49 & 29.13 & 0.55 & 0.01 & 0.03 & 0.08 & 100.18 & 1.1 & 77.9 & 21.0 \\
BEN-3       & 55.64 & 0.17 & 0.15 & 13.88 & 0.45 & 29.01 & 0.82 & 0.03 & 0.01 & 0.15 & 100.32 & 1.6 & 77.6 & 20.8 \\
BEN-4       & 55.68 & 0.22 & 0.11 & 14.12 & 0.50 & 29.13 & 0.49 & 0.01 & 0.03 & 0.11 & 100.40 & 0.9 & 77.9 & 21.2 \\
BEN-5       & 55.52 & 0.21 & 0.21 & 13.77 & 0.45 & 28.83 & 1.02 & 0.03 & 0.01 & 0.16 & 100.22 & 2.0 & 77.3 & 20.7 \\
BEN-6       & 55.69 & 0.17 & 0.16 & 14.05 & 0.50 & 29.03 & 0.77 & bdl  & 0.02 & 0.12 & 100.50 & 1.5 & 77.5 & 21.0 \\
BEN-7       & 55.79 & 0.17 & 0.17 & 13.58 & 0.50 & 29.04 & 0.75 & bdl  & 0.01 & 0.10 & 100.11 & 1.5 & 78.1 & 20.5 \\
BEN-8       & 55.69 & 0.16 & 0.13 & 14.07 & 0.50 & 29.22 & 0.69 & 0.01 & bdl  & 0.10 & 100.57 & 1.3 & 77.7 & 21.0 \\
BEN-9       & 55.77 & 0.19 & 0.17 & 13.99 & 0.48 & 28.86 & 0.89 & bdl  & 0.00 & 0.12 & 100.46 & 1.7 & 77.3 & 21.0 \\
BEN-10      & 55.70 & 0.21 & 0.15 & 13.73 & 0.51 & 28.89 & 0.70 & 0.01 & 0.02 & 0.12 & 100.04 & 1.4 & 77.9 & 20.8 \\
BEN-12      & 56.25 & 0.12 & 0.08 & 13.76 & 0.49 & 29.40 & 0.70 & bdl  & bdl  & 0.06 & 100.85 & 1.3 & 78.1 & 20.5 \\
BEN-13      & 55.96 & 0.11 & 0.10 & 13.97 & 0.51 & 29.26 & 0.78 & 0.01 & 0.02 & 0.06 & 100.78 & 1.5 & 77.7 & 20.8 \\
CHE-5       & 54.78 & 0.17 & 0.15 & 15.56 & 0.42 & 27.85 & 0.77 & bdl  & 0.02 & 0.12 & 99.82  & 1.5 & 75.0 & 23.5 \\
CHE-6       & 54.77 & 0.18 & 0.14 & 15.57 & 0.48 & 27.76 & 0.86 & 0.05 & 0.00 & 0.09 & 99.89  & 1.7 & 74.8 & 23.5 \\
CHE-7       & 54.86 & 0.15 & 0.16 & 15.91 & 0.43 & 27.58 & 0.70 & 0.02 & bdl  & 0.07 & 99.88  & 1.4 & 74.5 & 24.1 \\
CHE-8       & 54.90 & 0.14 & 0.11 & 15.49 & 0.46 & 27.84 & 0.84 & 0.00 & 0.04 & 0.09 & 99.90  & 1.6 & 75.0 & 23.4 \\
CHE-9       & 54.43 & 0.19 & 0.13 & 15.66 & 0.39 & 27.36 & 0.68 & 0.01 & 0.02 & 0.11 & 98.97  & 1.3 & 74.7 & 24.0 \\
CHE-10      & 54.96 & 0.16 & 0.11 & 15.57 & 0.41 & 27.58 & 0.64 & bdl  & 0.01 & 0.16 & 99.59  & 1.3 & 75.0 & 23.7 \\

\hline
\hline
\multicolumn{15}{l}{bdl: below detection limit} \\
\end{tabular}
\label{tab:2}
\end{table}
    
\section{Discussion} \label{sec:4}
\subsection{A source of low D/H ratios in Benenitra and Chelyabinsk} \label{subsec:4.1}
As shown in Fig. \ref{fig:4}, the $\delta$D$_{SMOW}$ values of orthopyroxene minerals in OC finds are distinguishable from those in OC falls, although there is some overlap. For clarity, we made the box-and-whisker plots (Fig. \ref{fig:5}), in which minerals in OC finds exhibit elevated interquartile ranges of both $\delta$D$_{SMOW}$ values and water contents relative to those of OC falls. Previous studies have revealed that Antarctic meteorites subjected to terrestrial weathering can cause localized mobilization of both major elements (e.g., Si and Mg, \citealp{velbel2014terrestrial}) and trace elements (e.g. rare earth elements, \citealp{crozaz2003chemical}). In addition, terrestrial exposure experiment conducted on the Martian meteorite fall Tissint showed that terrestrial weathering can rapidly affect the hydrogen isotopic ratios and water contents of olivines \citep{StephantAlice2018Teoa}. Therefore, long-term exposure in Antarctic ice has likely altered the original hydrogen signatures in small stones and caused the elevation of $\delta$D$_{SMOW}$ values and water contents in chondritic minerals. Furthermore, the Antarctic thin sections use tiny amounts of epoxy resin, which may be the likely reason for the elevated water contents and terrestrial-like D/H ratios in the OC falls \citep{mane2016hydrogen}. The contamination signal from the resin introduced during thin section preparation of the OC finds in this study cannot be completely avoided during SIMS measurements. Thus, Antarctic meteorite finds are not necessarily the best samples for hydrogen (and other volatile) analyses. Inferences about small Antarctic stones, especially thin sections prepared in epoxy should be made with caution.

Chelyabinsk and Benenitra were immediately collected and well preserved after they fell on Earth. Besides, the careful handling of the samples during polishing and mounting, and analysis of thick rock sections with no epoxy would keep terrestrial contamination to a minimal. Several parent body processes could have modified the original compositions of hydrogen in the minerals. These processes include (i) electron irradiation \citep[e.g.,][]{laurent2015deuterium}, (ii) galactic cosmic ray (GCR) spallation \citep[e.g.,][]{furi2017production}, (iii) heating events \citep[e.g.,][]{huss2006thermal}. Below, we discuss each of these processes and their effects on the hydrogen component. 

\textbf{\textit{(i) Electron irradiation}}. Electron irradiation can induce heating of the sample and ionization of hydrogen atoms from the surface of solids resulting in hydrogen loss coupled with deuterium enrichment of the solid residue \citep{laurent2015deuterium}. Thus, it is possible that D/H ratios in silicate minerals freely floating in the protosolar nebula prior to accretion or on the surface of the parent bodies are modified by electron irradiation. Previous experiments by \cite{roskosz2016experimental} reported that electrons in the keV regime have the capacity to produce large D-enrichment (fractionation of D/H can be up to 600$\permil$) in analogs of insoluble organic matter that is present in chondrites. However, the electron irradiation effects on silicate minerals are shown to be much smaller (fractionation of D/H is $<$200$\permil$). Particularly, fractionation of hydrogen isotopes caused by the electron irradiation in the silicate minerals larger than 10 $\mu$m is supposed to be $<$20$\permil$ \citep{roskosz2016experimental}. In our case, the measured orthopyroxenes in the OCs are all larger than 10 $\mu$m in size. More importantly, because these measured minerals are most likely the remnants of originally crystallized larger chondrules, the fractionation of D/H in the original large chondrules would be even smaller. Therefore, we argue for negligible modification of the hydrogen isotopic compositions caused by the electron irradiation. 

\textbf{\textit{(ii) Galactic Cosmic Ray spallation}}. Along with electron irradiation, spallation reactions triggered by the high energy galactic cosmic rays (GCRs) could occur at the planetesimal stage. The spallation reactions are responsible for the formation of secondary protons and generally limited to the surfaces of asteroid bodies (about 2–3 m) \citep{furi2017production}. The amount of the secondary particles produced in this mechanism are correlated with the exposure age and production rates. Using the reported exposure ages of OC meteorites (10–70 Ma, \citealp{marti1992cosmic}) and the production rates of hydrogen (4 × 10$^{-10}$ mol g$^{-1}$ Ma$^{-1}$, \citealp{merlivat1976spallation}) and deuterium (2.17 × 10$^{-12}$ mol g$^{-1}$ Ma$^{-1}$, \citealp{furi2017production}), we calculated the production rate of water. The results show small effects on both water contents and hydrogen isotopes: Less than 1 ppm water could have been produced by GCR spallation and the resulting hydrogen isotopic fractionation  ranges from 1$\permil$ to 26$\permil$ with an average value of 8$\permil$, which is within the analytical uncertainties (14–46$\permil$, 1$\sigma$). 

\textbf{\textit{(iii) Heating events}}. In OC parent bodies, heat can be produced by either decay of short-lived radionuclides, i.e., $^{26}$Al and $^{60}$Fe, which results in thermal metamorphism in planetesimals \citep{huss2006thermal}(see Section 4.2 for an extended discussion on thermal metamorphism), or impacts that cause localized high post-shock temperatures \citep{sharp2006shock}. In both these scenarios, high temperatures have been widely invoked to account for the dehydration of NAMs. Since hydrogen generally has higher diffusion rate than that of deuterium under same physio-chemical conditions, an elevation of D/H ratio is expected to exhibit along with the loss of water from NAMs. Additionally, a negative correlation between the D/H ratios and water contents of dehydrated minerals is expected after experiencing high-temperature events resulting in H loss. However, Chelyabinsk and Benenitra show no specific trend between the D/H ratios and water contents (Fig. \ref{fig:4}). Therefore, the variations of D/H ratios and water contents are not a result of dehydration during thermal metamorphism or impacts.

Rather than the dehydration of NAMs under high temperatures, NAMs could have received a fraction of water from the interstitial mesostases or matrix, as the latter components of unequilibrated chondrites can contain considerable amounts of water. Large variations in water contents of the chondrule mesostases from type 3 OCs (Bishunpur and Semarkona) have been reported (0.0008–1.51 wt.$\%$, Fig. \ref{fig:5}, \citealp{stephant2017water,shimizu2021highly}). The  $\delta$D$_{SMOW}$ values of this water range from -166$\permil$ to 15000$\permil$ \citep{stephant2017water,shimizu2021highly}. Besides,  $\delta$D$_{SMOW}$ values of the phyllosilicate in the matrix of the type 3 OC Semarkona have a large range and the upper limit can be up to $\sim$10000$\permil$ (Fig. \ref{fig:5}, \citealp{PianiLaurette2015MDhi}). Therefore, along with the re-crystallization of silicate components in the mesostases and matrix, varying degrees of mixing of water in both NAMs and the interstitial components would have resulted in the variable D/H ratios and water contents observed in Benenitra and Chelyabinsk NAMs. 

Nevertheless, several mineral grains in Benenitra and Chelyabinsk have much lower D/H ratios than the reported ones for the mesostases and matrix. This low D/H water reflects a D-poor source that has clearly been preserved in the NAMs, despite experiencing phase equilibration during thermal metamorphism. Because solar nebula is the only known reservoir characterized by a D-poor signature ($\delta$D$_{SMOW}$ $\approx$ -850 $\permil$), a portion of nebular hydrogen should have been incorporated as the primordial water into NAMs. Accordingly, we define this water as nebula-sourced water. The accretion of the chondrite parent bodies likely occured in short timescales. This requires a process that incorporates nebula-sourced water into the chondritic NAMs efficiently and quickly. Chondrules crystallized from their precursor melts in the gaseous nebula. It is possible that hydrogen gas directly dissolves into the precursor melts and subsequently is present in the chondrule crystals. However, the solubility of hydrogen in NAMs depends on the pressure subjected to the precursor melts. Petrologic experiments show that high pressures ($>$20 kbar) are required for NAMs to contain more than 200 ppm water \citep{mierdel2007water}. Thus, in the low-pressure nebular environment (10$^{-3}$ bar, \citealp{2000Sci...290.1751G}), hydrogen cannot be significantly introduced into the chondritic NAMs via an igneous process. Alternatively, hydrogen implantation is responsible for the incorporation of the nebula-sourced water \citep{jin2021hydration}. We then modelled the D/H variation observed in the measured NAMs assuming it were the result of the two-component mixing between the nebula-sourced D-poor water in the minerals and D-rich water in mesostases and matrix, based on the equation:
\begin{equation}\label{eq:3}
    \delta D_{MIX} = \delta D_{P} f + \delta D_{M} (1-f),
\end{equation}
where $\delta$D$_{MIX}$ is the measured $\delta$D$_{SMOW}$ value of the mineral from OC falls; $\delta$D$_P$ and $\delta$D$_M$ are the reported $\delta$D$_{SMOW}$ values of primordial nebular hydrogen (-850$\permil$, \citealp{geiss1998abundances}) and the water in mesostases or matrix of OCs (-166 – +10000$\permil$, \citealp{stephant2017water,PianiLaurette2015MDhi}); $f$ denotes the proportion of primordial water in the silicate minerals of OC falls. The $\delta$D$_M$ values and contents of water in the mesostases and matrix cover a large range. If we use the lowest reported $\delta$D$_{SMOW}$ value of the mesostasis (-166$\permil$), up to 45$\%$ of the water in the measured silicate minerals ($\delta$D$_{SMOW}$ = -400$\permil$) is the primordial nebula-sourced component. However, if we assume that $\delta$D$_{M}$ resembles that of the D-rich phyllosilicates ($\delta$D$_{SMOW}$ =$\sim$10000$\permil$), during the mixing, more than $95\%$ of the measured water is thus the primordial nebula-sourced water. Chondrule minerals may have received variable amounts of nebular-sourced water because of the stochastic nature of the implantation-diffusion process. Besides, the water contents and hydrogen isotope variations in the matrix is large. Thus, we argue that the observed variability in D/H ratios likely resulted during parent body thermal metamorphism.

To summarize, our results together with previously reported ones \citep[e.g.,][]{piani2020earth} argue that the water storage capability of chondritic pyroxene mineral is high. The water in OC pyroxenes is a mix of the D-poor nebula-sourced water and D-rich water from interstitial chondritic components. The low D/H ratios of some pyroxenes in OC falls Benenitra and Chelyabinsk indicate a large fraction (45-95$\%$) of the nebula-sourced water ($\delta$D$_{SMOW}$ = -850$\permil$ ). Electron irradiation and GCR spallation have barely modified the hydrogen component in the measured NAMs.   

\begin{figure}[ht]
    \centering
    \includegraphics[scale=0.15]{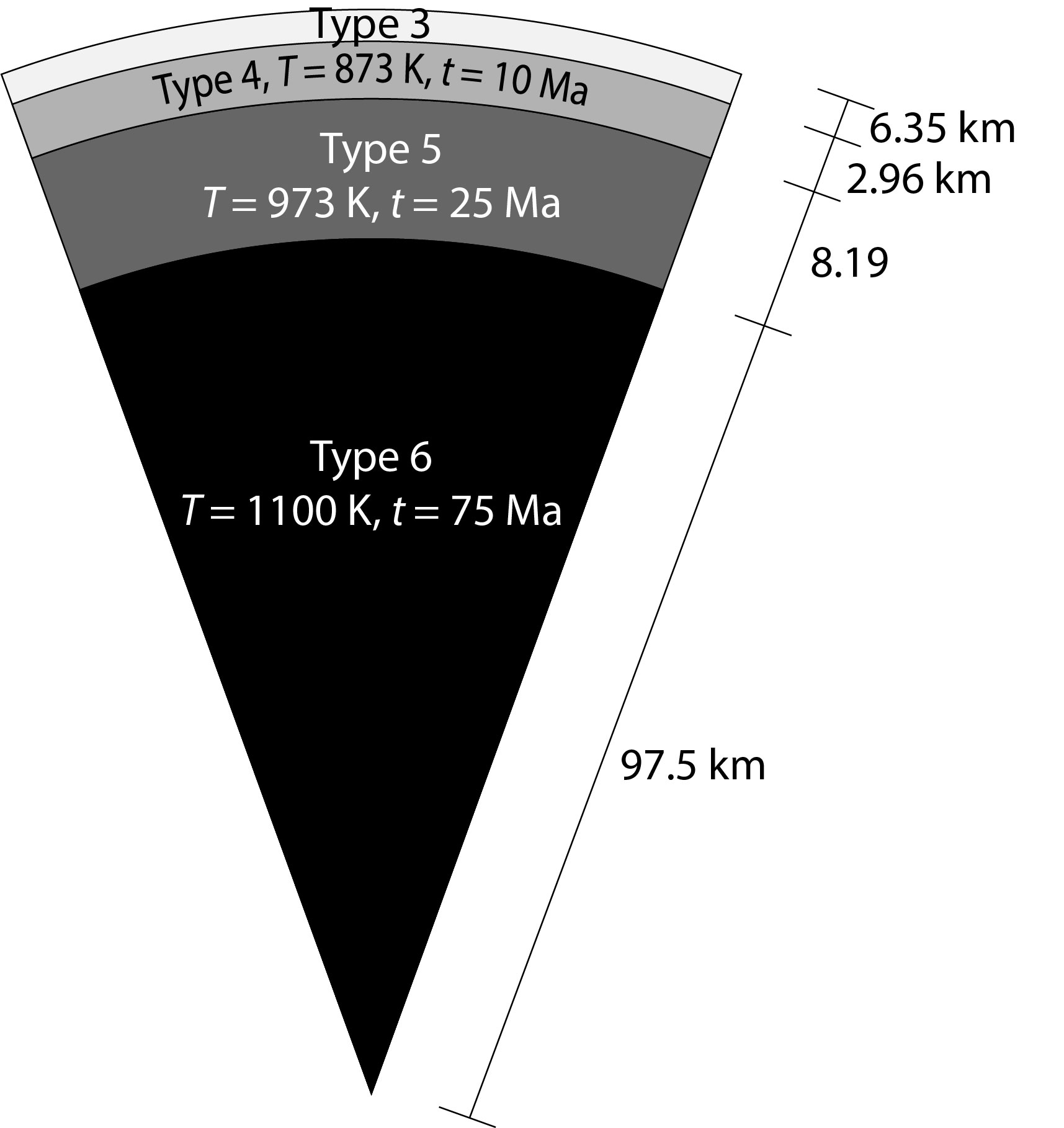}
    \caption{\textbf{Model of L (LL) group chondrite parent body used for the H diffusion simulations}. The thickness and peak temperatures of each type of material is from \cite{GailHans-Peter2019Thmo}. The duration ($t$) of the thermal metamorphism is adopted when the temperatures are higher than 600 K.}
    \label{fig:6}
\end{figure}

\subsection{The bulk water content of Benenitra and Chelyabinsk parent bodies}\label{subsec:4.2}

Considering the planetesimal as a whole, high temperatures have the ability to drive water/hydrogen out and reduce the bulk content of water. In a recent model proposed by \citep{GailHans-Peter2019Thmo}, the parent body of L/LL group of OCs is suggested to have a general radius of 115 km. From the center to the surface, the parent body is divided into type 6 ($>$17.5 km in depth), type 5 (9.3–17.5 km), type 4 (6.4–9.3 km), and type 3 ($<$6.4 km in depth) materials (Fig. \ref{fig:6}). The interior of the planetesimals have experienced prolonged metamorphism with higher peak temperatures, albeit with slow cooling rates. To estimate the amount of water that can be lost during thermal metamorphism in asteroid bodies, we built a hydrogen-diffusion model based on the corresponding solution to Fick’s second law for spherical geometry \citep{IngrinJ2006DoHi}: 
\begin{equation}\label{eq:4}
    F =  1-\frac{6}{\pi^2}\sum_{n=1}^{\infty} \frac{1}{n^2}\exp(\frac{-D_{eff} n^2 \pi^2 t}{d^2}),
\end{equation}
where $F$ is the fraction of the residual water/hydrogen in the materials after thermal metamorphism; $t$ (s) is the duration of the diffusion process caused by thermal metamorphism and adopted as 10 Ma, 25 Ma, and 75 Ma for petrologic types 4–6 materials; $d$ (m) is the distance of the minerals to the surface of the parent body; and $n$ denotes natural numbers. $D_{eff}$ is the effective diffusivity of the diffusing species in a porous parent body and is indicated by 
\begin{equation}\label{eq:5}
     D_{eff} = D_0 \exp(\frac{-\Delta H}{R_g T})\frac{\psi}{\tau},
\end{equation}
where $D_0$ (m$^2$ s$^{-1}$) is the diffusive coefficient for H in a specific material and $\Delta H$ (J mol$^{-1}$) is the activation enthalpy. $R_g$ is the gas constant (8.314 J mol$^{-1}$ K$^{-1}$) and $T$ (K) is the temperature at which diffusion took place. The porosity ($\psi$) of the parent body was assumed as 10$\%$ in our simulations, based on the analyses of 285 OCs \citep{1998M&PS...33.1221C}. The tortuosity ($\tau$) of the parent body were set as 1.5 according to the ultrasound measurements for four OCs \citep{ElAbassiD2013Asmf}. We searched the literature for published experimental data of $D_0$ and $\Delta H$ for olivine, diopside, and enstatite \citep{MackwellStephenJ1990Dohi,SunWei2019HIiS,StalderR2003Hdin,HerculeSarah1999HidD,DemouchyS2003Wdis,10.1093/petrology/egw058}, and calculated the $D_{eff}$ (6.7 × 10$^{-16}$–1.5 × 10$^{-10}$ m$^2$ s$^{-1}$) in a temperature range of 873–1100 K.

\begin{figure}[ht]
    \centering
    \includegraphics[scale=0.2]{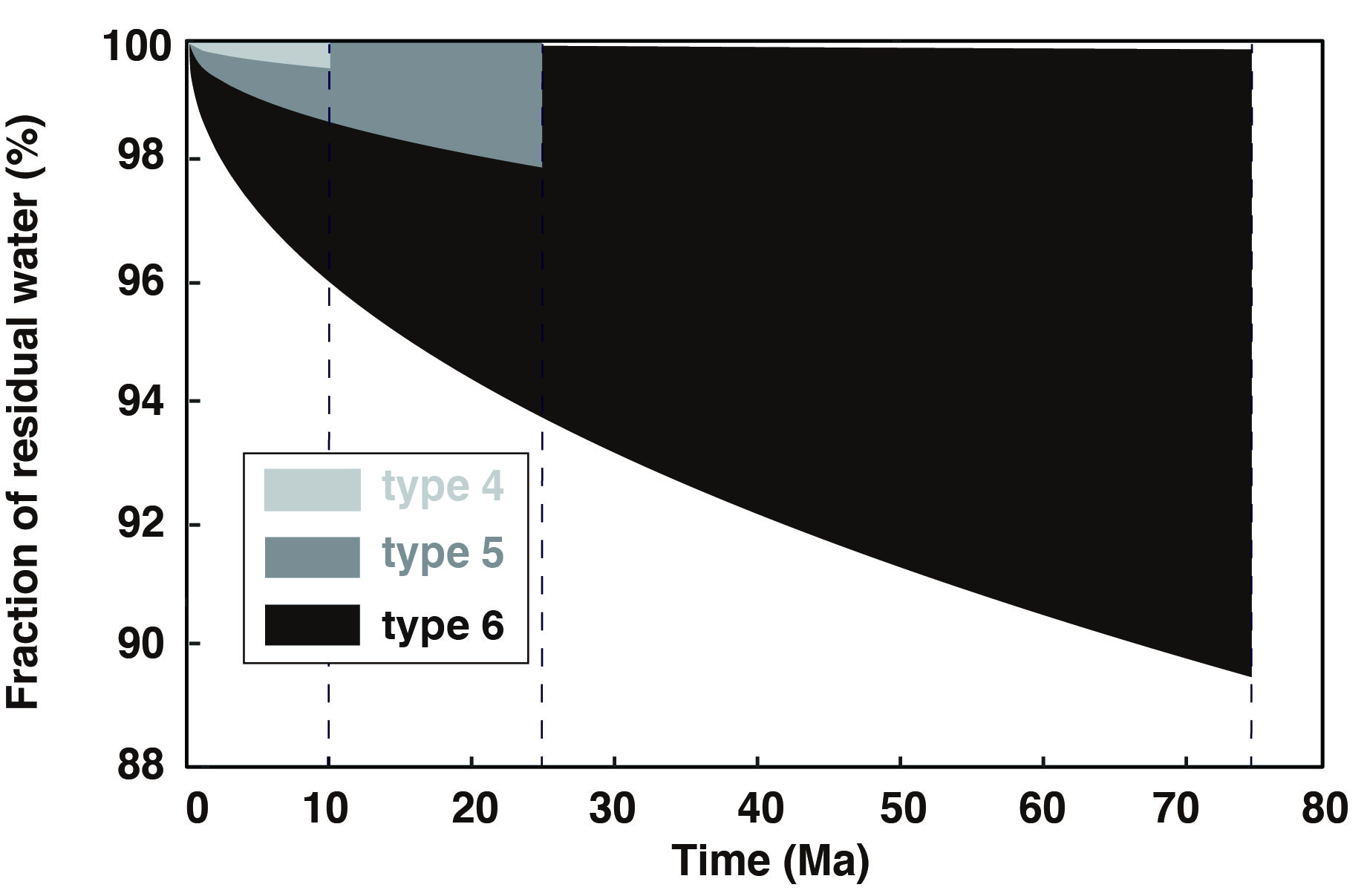}
    \caption{\textbf{Results of the thermal-diffusion model for OCs to quantify the fraction of residual water ($\%$) in silicate minerals after experiencing high temperatures}. The model is developed based on equations (\ref{eq:4}) and (\ref{eq:5}). The durations of the thermal metamorphism on types 4–6 are set as 10 Ma, 25 Ma, and 75 Ma; the peak temperature of the thermal metamorphism for type 4 materials is 873 K, type 5 is 973 K, and type 6 is 1100 K \citep{GailHans-Peter2019Thmo}. The fractions of the residual water (shaded areas) in the silicate minerals after thermal metamorphism in types 4 and 5 materials are higher than 98$\%$. Water loss from silicate minerals in type 6 materials is the largest but is still less than 12$\%$.}
    \label{fig:7}
\end{figure}

Our calculations show that the fraction of residual water in petrologic types 4 and 5 materials are above 98$\%$ (Fig. \ref{fig:7}). Type 6 material in the central regions of an undifferentiated body would lose $<$12$\%$ water, even with a thermal history of $\sim$75 Ma (Fig. \ref{fig:7}). We also simulated the same process based on another OC parent body model that proposes an 85 km-in-radius parent body with a somewhat different thermal profile \citep{miyamoto1982ordinary}, and observe that $<$10$\%$ water is lost during thermal metamorphism. In addition to the thermal metamorphism, heat can be locally produced by impacts and cause partial melting. Benenitra shows low weak shock stage (S3, \citealp{gibson2021investigation}) and Chelyabisk has been moderatly shocked (Stage S4, \citealp{ozawa2014jadeite}), indicating that their parent bodies have experienced impact events. However, because the cooling rates of ordinary chondritic materials are high (200-600K/thousand years, \citealp{ganguly2016cooling}), the high post-shock temperatures (up to 1500 K) could not last long to drive water/hydrogen out and modify the D/H ratios of minerals. Based on our model, in a time scale of 5 thousand years, the water content reduced by impacts is less than 1 ppm and the change of D/H ratios of minerals is smaller than 1$\permil$. Thus, our models clearly show that even at high temperatures (up to 900 $^\circ$C), parent bodies of Benenitra and Chelyabinsk (85–115 km in radius) would not lose significant amounts of water during thermal metamorphism.

Minerals from the studied Chelyabinsk and Benenitra meteorites resemble those from the S-type asteroid Itokawa, a parent body of type 4-6 ordinary chondrites \citep{nakamura2011itokawa}, we therefore estimated the bulk water content in OC parent bodies/S-type asteroids based on the analytical results of minerals from Benenitra, Chelyabinsk (this study) and Itokawa \citep{jin2019new,chan2021organic}. The basic assumptions in our calculations are: (1) The dominant components in primitive OC parent bodies are NAMs (olivine and pyroxene), metal grains, and matrix components. However, during thermal metamorphism, the matrix and mesostasis phases were transformed into feldspars.  (2) The average volume fraction of the chondrule minerals (olivine and pyroxene), feldspars, and metal grains in the metamorphized parent bodies are 85 vol.$\%$, 10 vol.$\%$, and 5 vol.$\%$, respectively. (3) Mineralogical fraction between olivine and pyroxene in S-type asteroids is pyroxene/(olivine+pyroxene) = 0.3–0.4. By adopting the following densities: 2.7 g cm$^{-3}$ for olivine, 3.3 g cm$^{-3}$ for pyroxene, 2.56 g cm$^{-3}$ for feldspar, and 7 g cm$^{-3}$ for metal, the mass fractions of 
olivine, pyroxene, feldspar, and metal grains in OC parent bodies are estimated as 48 wt.$\%$, 32 wt.$\%$, 8 wt.$\%$, and 11 wt.$\%$, respectively. Olivine and plagioclase mineral grains from Itokawa have been reported to contain 235±60 ppm and 993±252 ppm, respectively \citep{chan2021organic}. Water contents of pyroxene minerals measured in this study and those from Itokawa are in a range of 176-988 ppm \citep{jin2019new,chan2021organic}.  Using these numbers, our calculation shows that the structurally bound water in the NAMs can account for a mass fraction of 254-518 ppm in OC parent bodies or S-type asteroids.

\subsection{Water delivered by S-type asteroids to Earth and Mars}\label{subsec:4.3}

The estimated bulk water content of OC parent bodies here (254-518 ppm) is well in agreement with previously estimated water content of bulk silicate Itokawa parent body (160–510 ppm) and further confirms that inner solar system asteroids dominantly composed of NAMs can contain a significant amount of water. These hydrous small objects are the feeding material for planets; and via accretion, Earth and Mars received different amounts of water: the total water on Earth is $\sim$3900 ppm by weight \citep{peslier2017water}, whereas the bulk water content of Mars has been estimated as 83–320 ppm \citep{McCubbinFrancisM2019Oaao}. Here we evaluate the amount of water that could have been introduced by S-type asteroids to Earth and Mars, based on our analytical results and the existing planet accretion models. 

Earth formed inside the snow line from primarily inner solar system materials but may have acquired some fraction from beyond the snow line, especially late in its evolution \citep{wetherill1990formation,RaymondIzidoro2017}. In the Earth's accretion model proposed by \cite{DauphasNicolas2017Tino}, OC materials were suggested to account for 24 wt.$\%$ of Earth's accreting materials. Using this number and the calculated bulk water content in OC parent bodies (254-518 ppm), we estimate that water delivered by the S-type asteroids during Earth's accretion accounted for a mass fraction of $\sim$60–124 ppm weight, which corresponds to 0.3–0.7 times the equivalent mass of the oceans. This is comparable to the prediction by \cite{jin2019new} that $<$0.5 oceans water could have been delivered to Earth by accretion of S-type asteroids. In addition to S-type asteroids, enstatite chondrites can deliver 3 times the mass of Earth's oceans water \citep{piani2020earth}. 

The proportions of various chondritic materials accreted to form Mars are distinct from those for Earth. \cite{MahJingyi2021Idtp} recently developed the accretion model for terrestrial planets based on the isotopic data of meteorites as well as dynamic models of planet formation and suggested that S-type asteroids (H chondrites) would have contributed to up to 50$\%$ the mass of proto-Mars. The water brought by these S-type asteroids likely represents the proposed D-poor reservoir that existed in the mantle of Mars \citep{BoctorN.Z2003Tsow,BarnesJessicaJ2020Mewr}. Assuming parent bodies of H chondrites have an identical water content to that of the parent bodies of Chelyabinsk and Benenitra (254-518 ppm), the water delivered by S-type asteroids to Mars can account for a mass fraction of $\sim$127-259 ppm, which corresponds to more than half of the current bulk water content of Mars. Then, what is the source of the remaining water in Mars?

Although Earth and Mars formed concurrently in the inner solar system, the distinct bulk water content and the amounts of water delivered by one specific accreting materials likely reflect different accretion processes for these two planets. Currently, the two most widely discussed end-member accretion scenarios to form the terrestrial planets are planetesimal accretion or pebble accretion. In the scenario of planetesimal accretion, planetary embryos grow from the accretion of kilometre- to hundreds of kilometres-sized planetesimals in the early stage of planet formation \citep[e.g.,][]{1996Icar..124...62P,2005Icar..179..415H}. In contrast, pebble accretion involves the direct accretion of millimeter- to centimeter-sized dust grains that drift past the orbits of the growing protoplanets \citep[e.g.,][]{JohansenAnders2017FPvP,2019A&A...621C...1B}. An increasing number of studies reveal that both giant and terrestrial-sized planets could form through a combined mechanism that involves both pebble and planetesimal accretion \citep{AlibertYann2018TfoJ,LiuB2019Gats,2021Sci...371..365L}, however, the amount of the dominantly accreted solids may vary substantially from one protoplanet to another, sensitively depending on accretion chronology and the local pebble flux.

The relative contributions of pebbles and planetesimals not only influence the growth rate, but may affect the composition of the accreted material, which presents opportunities to further constrain the nature of the accreting material to the inner terrestrial planets. Inward drift of icy pebbles causes ice sublimation at the snow line resulting in loss of water \citep{2011ApJ...728...20S}; thus pebble-accreting embryos sample the local composition of pebbles. However, nebula-derived protoatmospheres of the planetary embryos would ablate the remaining water in the pebbles \citep{coleman2019pebbles}. The composition inherited from planetesimals depends both on their local material composition upon gravitational collapse by the streaming instability \citep{2017A&A...608A..92D,2021A&A...646A..14H} and subsequent compositional evolution before accretion onto growing protoplanets \citep{2015ApJ...804....9C,2019NatAs...3..307L}. Bulk elemental abundances of the martian core and mantle \citep{brennan2020core} indicate that the primordial material that formed Mars was significantly more oxidized than that of the Earth, which in the pebble accretion framework may be interpreted as a higher contribution of water-rich pebbles to the growing proto-Mars \citep{2016A&A...589A..15S}. However, the thermal requirements of core-mantle differentiation and bulk volatile fraction substantially limit the maximum amount of oxidized pebbles that can contribute to the growth of Mars \citep{2021JGRE..12606754Z}. In a hybrid pebble-planetesimal accretion framework, the increased oxidation state of Mars-forming materials relative to Earth can be explained by higher contributions from early-formed, oxidized planetesimals in the inner Solar System that seed the growth of the terrestrial planets \citep{2021Sci...371..365L}, or secondary, minor addition of pebbles to the Mars-forming embryos when the snowline migrates inward during later disk stages \citep{2011ApJ...738..141O,2016A&A...589A..15S}. Further work aiming to constrain the nature of the planet-forming material in the inner Solar System will be aided by the enhanced constraints on the bulk water fraction in meteoritic materials that we derived here.

\section{Conclusions} \label{sec:5}
(1) While the Antarctic OC finds show clear evidence of terrestrial contamination leading to elevated D/H ratios and water contents, the low D/H ratios detected in NAMs from fresh OC falls, Chelyabinsk and Benenitra exhibit a component of primordial D-poor nebular hydrogen. 

(2) The variations of measured D/H ratio and water contents in Chelyabinsk and Benenitra are likely a result of varying degrees of water mixing between NAMs and hydrous interstitial chondritic components during parent body processes.

(3)  Water-loss in S-type asteroids during thermal metamorphism is minimal. The bulk of the water in the parent bodies of Chelyabinsk and Benenitra should survive this process, even after reaching temperatures of $\sim$800$^\circ$C and is reported to 254-518 ppm. These water contents are consistent with the estimated water content of Itokawa parent body. 

(4) S-type asteroids were a potential source of water on both Earth and Mars. Proto-Earth received water equivalent to 0.3–0.7 Earth's oceans primarily through accretion of S-type asteroids while proto-Mars gained $\sim$127-259 ppm water. The existence of the nebular water component in silicate minerals \citep{jin2021hydration} and their incorporation into inner solar system objects reduces the need for a late water delivery to the terrestrial planets.

(5) Mars may have have received most of its water from accretion of early-formed, oxidized planetesimals, with a minor addition of icy pebbles to the Mars-forming embryos, when the snowline migrated inward during later disk phases.

\acknowledgments

We thank Philipp Heck at the Field Museum for providing fresh fragments of ordinary chondrite falls, Johnson Space Center for providing thin sections ordinary chondrite finds, and Smithsonian Institution and Kathryn. M. Kumamoto at Stanford University for the standard samples. We would like to profusely thank the anonymous reviewers for providing constructive comments and suggestions that helped improve the manuscript. Axel Wittmann and Richard L. Hervig (NSF EAR grant 1819550) are thanked for their assistance with electron probe and IMS 6f operations at ASU. Z. J. was funded by the startup funds from the State Key lab of Lunar and Planetary Science of Macao University of Science and Technology. M. B. was funded by her startup funds from ASU for the analytical work. T. L. was supported by a grant from the Simons Foundation (SCOL award No. 611576). 
G. D. M. acknowledges support from ANID --- Millennium Science Initiative ---  ICN12\_009.
This material is based upon work supported by the National Aeronautics and Space Administration under Agreement No. 80NSSC21K0593 for the program “Alien Earths”. The results reported herein benefitted from collaborations and/or information exchange within NASA’s Nexus for Exoplanet System Science (NExSS) research coordination network sponsored by NASA’s Science Mission Directorate.

\bibliography{Hydrogen}
\bibliographystyle{aasjournal}
\end{document}